\documentclass[twocolumn,nofootinbib]{revtex4-2}

\usepackage{float}
\usepackage[dvipsnames]{xcolor}
\usepackage{graphicx}
\usepackage{dcolumn}
\usepackage{bm}
\usepackage{amsmath}
\usepackage{hyperref}
\usepackage{subfigure}
\usepackage{verbatim}
\usepackage{amssymb}
\usepackage{ulem}

\hypersetup{
   citebordercolor=CornflowerBlue,
   urlbordercolor=SpringGreen,
  linkbordercolor=Apricot
}

\newcommand{\Planck}{\textit{Planck}}
\newcommand{\hhat}[1]{\hat{\hat{#1}}}
\newcommand{\tot}{\mathrm{tot}}

\begin{document}
\title{Revisiting the impact of neutrino mass hierarchies on neutrino mass constraints \\
in light of recent DESI data}

\author{Laura Herold}
\email{lherold@jhu.edu}
\author{Marc Kamionkowski}
\email{kamion@jhu.edu}

\affiliation{\vspace{2mm} William H. Miller III Department of Physics and Astronomy, Johns Hopkins University, 3400 North Charles Street, Baltimore, Maryland 21218, USA}

\begin{abstract}
Recent results from DESI combined with cosmic microwave background data give the tightest constraints on the sum of neutrino masses to date. However, these analyses approximate the neutrino mass hierarchy by three degenerate-mass (DM) neutrinos, instead of the normal (NH) and inverted hierarchies (IH) informed by terrestrial neutrino oscillation experiments. Given the stringency of the upper limits from DESI data, we test explicitly whether the inferred neutrino mass constraints are robust to the choice of neutrino mass ordering using both Bayesian and frequentist methods. 
For \Planck\ data alone, we find that the DM hierarchy presents a good approximation to the physically motivated hierarchies while showing a strong dependence on the assumed lower bound of the prior, confirming previous studies. 
For the combined \Planck\ and DESI baryon acoustic oscillation data, we find that assuming NH ($M_\tot < 0.13\,\mathrm{eV}$) or IH ($M_\tot < 0.16\,\mathrm{eV}$) loosens the Bayesian upper limits compared to the DM approximation ($M_\tot < 0.086\,\mathrm{eV}$).
The frequentist analysis shows that the different neutrino models fit the data equally well and the loosening of the constraints can thus be attributed to the lower bounds induced by NH and IH. 
Overall, we find that the DM hierarchy presents a good approximation to the physically motivated hierarchies also for \Planck+DESI data as long as the corresponding lower neutrino mass bounds are imposed.
\end{abstract}

\maketitle


\section{Introduction}
\label{sec:intro}

The Dark Energy Spectroscopic Instrument (DESI) collaboration has recently reported the tightest cosmological neutrino mass bounds to date \cite{DESI:2024mwx, DESI:2024aax, DESI:2024uvr, DESI:2024lzq}, inferring an upper limit on the sum of neutrino masses of $M_\tot < 0.082 \,\mathrm{eV}$\footnote{Due to an update in the ACT lensing likelihood, the previous tighter constraint, $M_\tot < 0.072 \,\mathrm{eV}$, from \Planck+ACT+DESI BAO data reported in \cite{DESI:2024mwx} loosened to $M_\tot < 0.082 \,\mathrm{eV}$ as pointed out in \cite{Jiang:2024viw, DESI:2024hhd}.} \cite{DESI:2024hhd} when combining DESI baryon acoustic oscillation (BAO) data with cosmic microwave background (CMB) data from the \Planck\ experiment \cite{Planck:2019nip, Planck:2018vyg} and the Atacama Cosmology Telescope (ACT, \cite{ACT:2023dou, ACT:2023kun}). This tightens further when combining these data with the DESI measurement of the full shape of the galaxy power spectrum, $M_\tot < 0.071 \,\mathrm{eV}$ \cite{DESI:2024hhd}.\footnote{We quote all upper limits in this paper at $95\%$ confidence limit if not otherwise indicated.}

These tight neutrino mass constraints begin to challenge the lower limits inferred from terrestrial neutrino oscillation experiments. Neutrino oscillation experiments are not directly sensitive to the individual neutrino masses, $m_1,\, m_2,\, m_3$, but only to the absolute mass differences between the three neutrino species. This does not uniquely determine the hierarchy (or ordering) of the masses and one is left with two options: the normal hierarchy (NH) with $m_1 < m_2 \ll m_3$  and the inverted hierarchy (IH) with $m_3 \ll m_1 < m_2$. The total sum of neutrino masses, $M_\tot = m_1 + m_2 + m_3$, can then be computed as:
\begin{equation*}
\resizebox{1.\hsize}{!}{$
    \begin{split}
    M_\tot &= m_0 + \sqrt{\Delta m^2_{21} + m_0^2} + \sqrt{\Delta m_{31}^2 + m_0^2}, \hspace{2cm} \mathrm{(NH)}\\
    M_\tot &= m_0 + \sqrt{|\Delta m_{32}^2| + m_0^2} + \sqrt{|\Delta m_{32}^2| - \Delta m^2_{21} + m_0^2}, \quad \mathrm{(IH)}
    \end{split}
$}
\end{equation*}
with $m_0 = m_1$ for NH and $m_0 = m_3$ for IH. A combined fit to different oscillation experiments gives for the mass splittings \cite{Esteban:2016qun} (or \cite{Gonzalez-Garcia:2021dve}): $\Delta m^2_{21} = m_2^2 - m_1^2 = 7.5_{-0.17}^{+0.19}\cdot 10^{-5}\,\mathrm{eV}^2$ and $|\Delta m^2_{3\ell}| = |m_3^2 - m_\ell^2| = 2.45_{-0.04}^{+0.04}\cdot 10^{-3}\,\mathrm{eV}^2$, where $\ell = 1$ for NH and $\ell = 2$ for IH. Setting the smallest neutrino mass to zero, $m_0 = 0$, and inserting $\Delta m^2_{21}$ and $\Delta m^2_{3\ell}$, allows to determine the minimum sum of neutrino masses in the two hierarchies:
\begin{equation}
\label{eq:min_mass}
\begin{split}
    & M_\tot \gtrsim 0.06\, \mathrm{eV} \quad \mathrm{(NH)}, \\
    & M_\tot \gtrsim 0.1\, \mathrm{eV}\ \ \quad \mathrm{(IH).}
\end{split}
\end{equation}
While there is no strong preference for either of the neutrino hierarchies from neutrino oscillation data \cite{Esteban:2020cvm, deSalas:2020pgw, Capozzi:2021fjo} and upper limits from terrestrial measurements are still weak, e.g. $m_\nu \lesssim 0.45\, \mathrm{eV}$ (90\% C.L.) for the effective electron-neutrino mass from the KATRIN experiment \cite{Katrin:2024tvg}, the tight upper limits from cosmological data, in particular the recent DESI results, seem to be increasingly in favor of the NH (see e.g.\ \cite{Hannestad:2016fog, Gerbino:2016ehw, Vagnozzi:2017ovm, Simpson:2017qvj, Schwetz:2017fey, Gariazzo:2018pei, Jimenez:2022dkn, Gariazzo:2023joe} for discussions about evidence for either hierarchy in cosmology). This preference for very small, vanishing or even ``negative'' neutrino masses has been studied extensively in the literature (e.g.\ \cite{Craig:2024tky, Green:2024xbb, Noriega:2024lzo, Naredo-Tuero:2024sgf, Elbers:2024sha}).

However, the current DESI results \cite{DESI:2024mwx, DESI:2024hhd} -- like most studies -- approximate the physically motivated NH and IH by  assuming that the three species of massive neutrinos each carry one third of the total mass: $m_1 = m_2 = m_3 = M_\tot /3$, which we refer to as the ``degenerate-mass hierarchy'' (DM). This approximation has been shown \cite{Lesgourgues:2004ps, Lesgourgues:2006nd, Hannestad:2016fog, RoyChoudhury:2019hls} to be accurate for WMAP \cite{WMAP:2012fli} and \Planck\ data. Even optimistic forecasts, including experiments like CMB-S4 \cite{Abazajian:2019eic}, LiteBIRD \cite{Matsumura:2013aja}, Euclid \cite{EUCLID:2011zbd} and SKA \cite{Maartens:2015mra} seem to be unable to distinguish between the different hierarchies \cite{Archidiacono:2020dvx}. Ref.~\cite{CORE:2016npo} finds that using the DM approximation with a lower limit informed from NH or IH (Eq.~\ref{eq:min_mass}) gives constraints in good agreement with a full NH or IH analysis in forecasts for the CORE experiment.\footnote{Ref.~\cite{Wagner:2012sw} studies non-linear effects induced by the explicit neutrino mass ordering that could help to distinguish these. Moreover, future weak lensing \cite{DeBernardis:2009di, Jimenez:2022dkn} or line-intensity mapping surveys\cite{Pritchard:2008wy, Bernal:2021ylz} could be sensitive to possible neutrino decay channels.} Another common approximation is to assume one massive species and two massless ones, $m_1=M_\tot,\ m_2 = m_3 = 0$, which we denote as ``1 massive / 2 massless'' (1M). Although commonly used in the literature, the 1M hierarchy has been shown to lead to a background evolution and matter power spectrum that are significantly different form NH, IH and DM \cite{Hannestad:2003ye, Lesgourgues:2004ps, Crotty:2004gm, Lesgourgues:2012uu, Lesgourgues:2013}.

While these studies indicate that the DM approximation presents a good approximation for cosmological data, we deem it important to confirm this explicitly for recent DESI BAO data given the stringency of the neutrino mass constraints as well as the possibility that the true neutrino masses may differ from the canonical values used in some of the earlier forecasting work. Hence, the goal of this paper is to check whether the inferred tight upper limits on $M_\tot$ are robust to the choice of neutrino mass ordering. To do so, we infer $M_\tot$ within both Bayesian methods based on MCMC posteriors and frequentist methods based on profile likelihoods, while considering the DM and 1M approximations and the physically motivated NH and IH. 
The paper is structured as follows; we give a short review about the effect of massive neutrinos on cosmological observables in Sec.~\ref{sec:neutrino_physics}, we describe the data sets and Bayesian/frequentist methodology in Sec.~\ref{sec:data_methods}, we discuss our results in Sec.~\ref{sec:results} and conclude in Sec.~\ref{sec:conclusions}.


\section{Massive neutrino's impact on cosmology}
\label{sec:neutrino_physics}

In this section, we briefly review massive neutrinos' effects on the geometry and the growth of structure of the universe following \cite{Lesgourgues:2013, Loverde:2024nfi}  (see also \cite{Lesgourgues:2006nd, Wong:2011ip, Lesgourgues:2012uu, Archidiacono:2016lnv}), where we focus on the impact on the data sets under consideration in this work (CMB power spectra, CMB lensing and BAO). 

The geometry of the universe is affected by the neutrinos' transition from relativistic to non-relativistic particles at a redshift depending on their total mass, $M_\tot$. Even for the most conservative bounds, massive neutrinos become non-relativistic only after recombination, since otherwise the early Integrated Sachs-Wolfe effect would be affected at an unacceptable level \cite{Planck:2018vyg, Hou:2012xq}. Hence, massive neutrinos contribute to the radiation fraction, $\Omega_r$, before recombination, while after recombination, they contribute to the matter fraction, $\Omega_m$, on scales larger than the free-streaming scale ($k<k_\mathrm{fs}$) and to $\Omega_r$ on scales smaller than the free-streaming scale ($k>k_\mathrm{fs}$). Since $\Omega_m(z) \sim (1+z)^3$ redshifts slower than $\Omega_r(z) \sim (1+z)^4$, this increases the expansion rate, $H(z) = H_0 \sqrt{\Omega_r (1+z)^4 + \Omega_m (1+z)^3 + \Omega_\Lambda}$, in the late universe compared to a universe with massless neutrinos. Taking all other parameters as fixed, this would decrease the angular diameter distance to last scattering, $D_A = \int_0^{z^*} \mathrm{d} z / H(z)$. Since however, the angular size of the sound horizon, $\theta_s = r_s/D_A$, is precisely measured by the CMB \cite{Planck:2018vyg}, where $r_s$ is the physical size of the sound horizon, the relative increase in $\Omega_m$ needs to be compensated by a lower $H_0$, leading to the well-known negative $M_\nu$-$H_0$-degeneracy. This degeneracy can be partially broken by including BAO data, which is sensitive to both $H(z)$ and $\Omega_m(z)$ at a redshift $z$. Thus all additional $\Omega_m$ present at the BAO redshift $z$ compared to the CMB-inferred $\Omega_m$ can be attributed to massive neutrinos. 

Massive neutrinos not only affect the geometry, but also the growth of structure of the universe: The excess energy density of massive neutrinos, and thus larger $H(z)$, leads to a larger Hubble friction, which slows down the growth of perturbations on all scales. On scales larger than $k_\mathrm{fs}$, this effect is compensated by the relative increase in $\Omega_m$ that leads to increased clustering through gravity. 
Neutrino free streaming, i.e.\ the neutrinos' inability to cluster at $k>k_\mathrm{fs}$, suppresses the matter power spectrum at scales smaller than  $k_\mathrm{fs}$. This suppression of structure growth results in a suppressed CMB lensing signal.

Both effects discussed above are dominated by the total mass, $M_\tot$. The distribution of $M_\tot$ among the three neutrino species could, in principle, impact cosmological observables, but the effects are small as shown by \cite{Archidiacono:2020dvx, Xu:2020fyg}. We briefly review their results here, comparing the NH/IH scenarios with the DM approximation first, and commenting on 1M in the end. The NH and IH have one ($m_3$) or two neutrinos ($m_1,\, m_2$), respectively, that are more massive than the three neutrinos in the DM approximation and thus become non-relativistic earlier. Since matter redshifts slower than radiation, this leads to excess energy density in NH/IH compared to DM at redshifts $z\sim 1-100$. As described above, excess energy leads a decrease in $D_A$ and thus to small shifts of $\theta_s$, i.e.\ small (sub-cosmic-variance) shifts of the acoustic peaks in the CMB power spectrum and the BAO scale. 

The excess energy density in NH/IH compared to DM impacts the growth of structure similar to the description above:  on scales larger than $k_\mathrm{fs}$, the increased $H(z)$ leads to a suppressed growth of perturbations. Moreover, the scale $k_\mathrm{fs}$ itself experiences small shifts due to the different mass orderings: $k_\mathrm{fs}$ is dominated by the mass of the heaviest neutrino, which is larger in NH/IH than in DM. Hence free streaming occurs on smaller scales in NH/IH than in DM. These two effects lead to an overall suppression of CMB lensing in NH/IH compared to DM. 

The 1M approximation features one neutrino that is even heavier and becomes non-relativistic even earlier than in the NH/IH scenarios, leading to even more excess energy. Hence, the shifts in $\theta_s$ are larger, the growth of perturbations is even more suppressed and the free streaming occurs at even smaller scales than in NH/IH/DM \cite{Archidiacono:2020dvx}.


\section{Data sets and methodology}
\label{sec:data_methods}

We consider the full DESI BAO sample \cite{DESI:2024mwx} (referred to as DESI), which measures the angular diameter distances $D_A(z)/r_\mathrm{d}$ and Hubble distances $D_H(z)/r_\mathrm{d}$ relative to the sound horizon at baryon drag, $r_\mathrm{d}$, in seven redshift bins from different tracers, taking into account correlations between $D_A(z)/r_\mathrm{d}$ and $D_H(z)/r_\mathrm{d}$.\footnote{The likelihood compatible with the \texttt{Cobaya} sampler is publicly available at \url{https://github.com/cosmodesi/desilike}. We adapt this likelihood to be compatible with the \texttt{MontePython} sampler. This likelihood and notebooks to reproduce the plots are available at \url{https://github.com/LauraHerold/MontePython_desilike}.}.  
We combine the BAO data with CMB data from \Planck\ PR3 TT, TE, EE and lensing power spectra \cite{Planck:2019nip, Planck:2018vyg} (referred to as \Planck), which calibrates above $r_\mathrm{d}$. 
The DESI BAO 2024 baseline results \cite{DESI:2024mwx} additionally include CMB lensing measurements from \Planck\ PR4 \cite{Carron:2022eyg} and ACT \cite{ACT:2023kun}. However, including these data requires an increase in precision in order to get accurate predictions at non-linear scales, which slows down the computations significantly. This combined with the increased complexity of the neutrino hierarchy modeling led to prohibitively slow parameter inference. Moreover, the cosmological constraints from \Planck+ACT+DESI show a (small) dependence on the non-linear modeling \cite{Jiang:2024viw} and the precision settings need to be chosen with care to get accurate enough predictions at small scales \cite{McCarthy:2021lfp}. Hence, we exclude ACT and \Planck\ PR4 data from our analysis but expect that it will not alter the main conclusions of this work. 

We use the Boltzmann solver \texttt{CLASS} \cite{Lesgourgues:2011re, Blas:2011rf} for predictions of the linear spectra assuming the $\Lambda$CDM model and \texttt{halofit} for non-linear corrections \cite{Smith:2002dz, Takahashi:2012em}. 
For all computations in this paper, we fix the number of relativistic degrees of freedom, $N_\mathrm{eff} = 3.046$.
We adopt both a Bayesian and frequentist analysis. For the Bayesian analysis, we compute posteriors with the MCMC sampler \texttt{MontePython} \cite{Audren:2012wb, Brinckmann:2018cvx}. We sample $M_\tot$, while the individual neutrino masses are computed within \texttt{MontePython} using the neutrino oscillation data from \cite{Esteban:2016qun} and passed to \texttt{CLASS}. We consider the chains as converged when a Gelman-Rubin criterion of $R-1 < 0.01$ is reached. For plotting the posteriors, we use \texttt{getdist} \cite{Lewis:2019xzd} allowing for adaptive smoothing. 

For the frequentist analysis, we compute profile likelihoods with \texttt{pinc} \cite{Herold:2024enb}\footnote{\url{https://github.com/LauraHerold/pinc}}. In a nutshell, a profile likelihood in a parameter of interest, here $M_\tot$, is given by
\begin{equation}
    \Delta\chi^2 (M_\tot) = -2\log \left(\frac{\mathcal{L}(M_\tot,\hhat{\boldsymbol{\nu}})}{\mathcal{L}(\hat{M}_\tot,\hat{\boldsymbol{\nu}})}\right),
\end{equation}
where $\mathcal{L}$ is the likelihood, $\boldsymbol{\nu}$ denotes all remaining cosmology and nuisance parameters, $\hhat{\boldsymbol{\nu}}$ is the conditional maximum likelihood of $\boldsymbol{\nu}$ for a fixed value of $M_\tot$ and $(\hat{M}_\tot,\hat{\boldsymbol{\nu}})$ is the global maximum likelihood or ``bestfit''. In other words, the profile likelihood is obtained by minimizing or ``profiling'' the $\Delta\chi^2$ over all parameters $\boldsymbol{\nu}$ for a fixed value of $M_\tot$, where for a Gaussian, $\Delta\chi^2$ indeed follows a $\chi^2$-distribution. 
Profile likelihoods, although more common in particle physics, have been extensively used to infer neutrino constraints from cosmology \citep{Reid:2009nq, Gonzalez-Morales:2011tyq, Planck:2013nga, Couchot:2017pvz, Reeves:2022aoi, Naredo-Tuero:2024sgf, Herold:2024enb} since these are known to have a strong dependence on the prior \cite{Simpson:2017qvj, Gariazzo:2018pei, Gariazzo:2023joe}.

Since the parameter of interest in this work, $M_\tot$, is close to its physical boundary in $M_\tot = 0$, we make use of the boundary-corrected graphical construction, also known as Feldman-Cousins construction \cite{Neyman:1937uhy, Feldman:1997qc} (for a detailed description of this construction in a cosmology context see \cite{Herold:2024enb}). 
This construction relies on the asymptotic assumptions such as normality, Wald's and Wilks' theorems \cite{Wilks:1938dza, Wald:1943}. These assumptions have been explored for $M_\tot$ within the DM approximation and have been found to be consistent with the asymptotic assumptions for \Planck-lite data \cite{Herold:2024enb}. Here, we do not explicitly test these assumptions and acknowledge that frequentist coverage might only be approximately given.


\section{Results}
\label{sec:results}

\begin{figure*}
    \centering
    \includegraphics[width=1\linewidth]{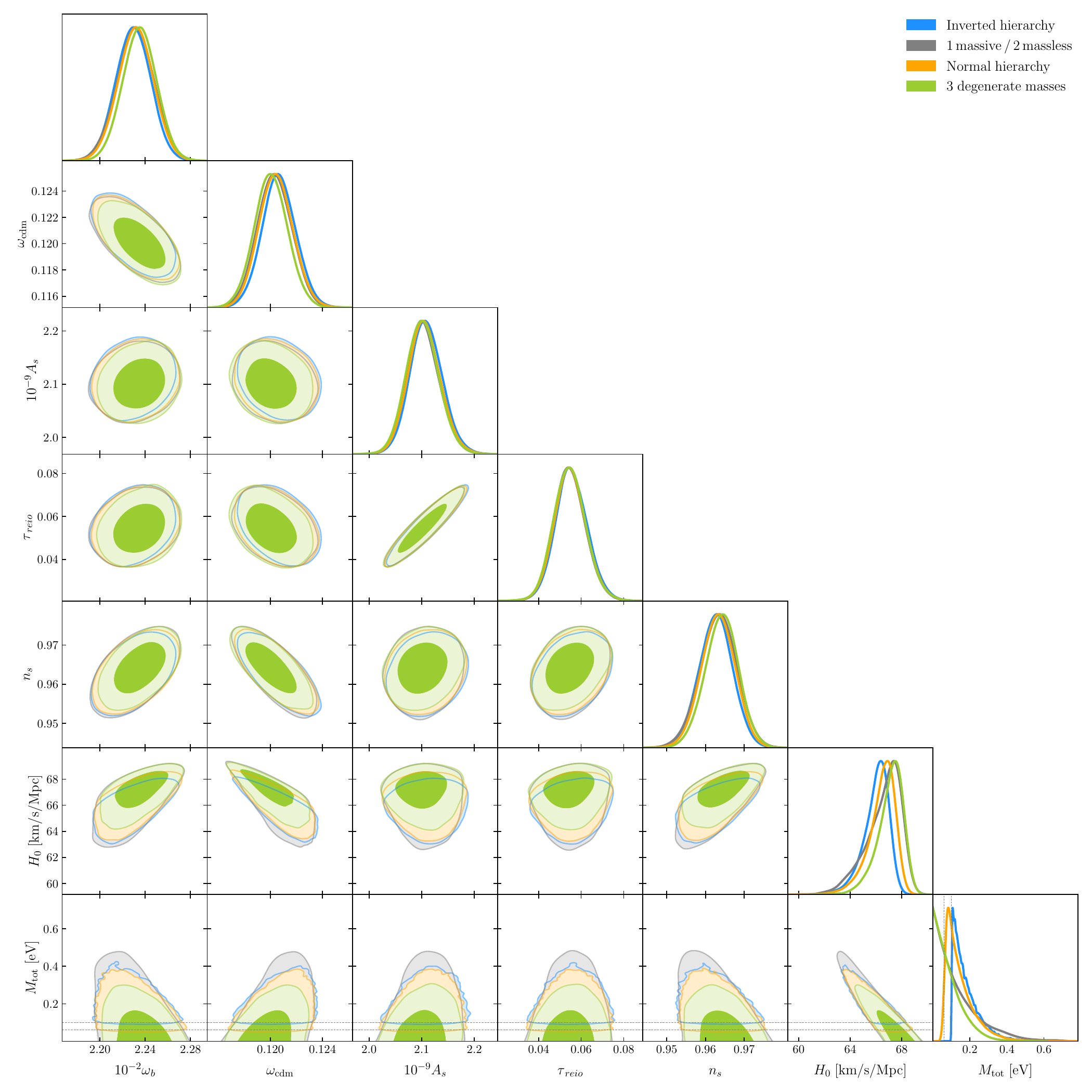}
    \caption{Posterior corner plot under \textbf{\Planck\ data} for different neutrino mass orderings. The 3-degenerate-masses approximation gives the tightest constraints on $M_\tot$, which loosen when considering the physically motivated normal and inverted hierarchies due to the lower limit of the prior imposed by neutrino oscillation experiments (dotted lines in $M_\tot$ panels).}
    \label{fig:MCMC_Planck}
\end{figure*}
To explore the impact of the neutrino mass ordering on the upper limits in $M_\tot$ inferred from cosmology, we first revisit constraints from \Planck\ CMB data alone, followed by \Planck+DESI data.

\subsection{\Planck}

\begin{table}
    \centering
    \begin{tabular}{c|c|c}
         Hierarchy & Bayesian        & Frequentist (NH/IH lower limit) \\
         \hline
         DM        & $< 0.24\,\mathrm{eV}$ & $< 0.18\ (0.23/0.26)\,\mathrm{eV}$ \\
         1M        & $< 0.39\,\mathrm{eV}$ & $< 0.20\ (0.25/0.29)\,\mathrm{eV}$ \\
         NH        & $< 0.31\,\mathrm{eV}$ & $< 0.16\ (0.22/0.25)\,\mathrm{eV}$ \\
         IH        & $< 0.34\,\mathrm{eV}$ & $< 0.15\ (0.20/0.25)\,\mathrm{eV}$ \\
    \end{tabular}
    \caption{95\% upper limits on the sum of neutrino masses, $M_\tot$, under \textbf{\Planck\ data} for different neutrino mass hierarchies. For the frequentist interval construction, we assumed $M_\tot >0$ ($M_\tot >0.06$/$0.1\,\mathrm{eV}$) as a physical lower boundary (see text).}
    \label{tab:Planck_constraints}
\end{table}
The results of the Bayesian analysis under \Planck\ CMB data are summarized in the left column of Tab.~\ref{tab:Planck_constraints} and the posterior corner plot is shown in Fig.~\ref{fig:MCMC_Planck}. For the most commonly used approximation, which assumes three neutrinos with degenerate masses (DM), we find $M_\tot < 0.24\, \mathrm{eV}$ in agreement with the \Planck\ collaborations' 2018 result \cite{Planck:2018vyg}. This constraint loosens to $M_\tot < 0.39\, \mathrm{eV}$ when adopting the alternative commonly used approximation of one massive neutrino and two massless ones (1M). As mentioned above, the 1M approximation leads to significant changes in cosmological observables compared to NH/IH/DM; here, we consider it regardless for comparison. Next, we adopt the physical normal (NH) and inverted hierarchies (IH) with mass splittings informed from neutrino oscillation experiments~\cite{Esteban:2016qun}, which lead to a physical lower limit of the sum of neutrino masses (Eq.~\ref{eq:min_mass}, dotted lines in Fig.~\ref{fig:MCMC_Planck}). As is evident in the bottom right panel of Fig.~\ref{fig:MCMC_Planck}, these lower limits cut off the posterior at higher $M_\tot$ leading to loosened constraints compared the DM approximation: $M_\tot < 0.31\,\mathrm{eV}$ (NH) and $M_\tot < 0.34\,\mathrm{eV}$ (IH). Hence, for the constraints under \Planck\ data alone, assuming the physically motivated NH and IH leads to a loosening of the constraints compared to the DM assumption, driven by imposing a lower limit, re-confirming the findings in \cite{Lesgourgues:2004ps, Lesgourgues:2006nd, Couchot:2017pvz, RoyChoudhury:2019hls}. We explore the impact of the prior further with profile likelihoods below.

The posteriors of the other $\Lambda$CDM parameters like the baryon and cold dark matter fractions ($\omega_b$, $\omega_\mathrm{cdm}$), the optical depth ($\tau$, not shown), and the amplitude and spectral index of the primordial power spectrum ($A_s$, $n_s$) in Fig.~\ref{fig:MCMC_Planck} show only negligible shifts for the different neutrino mass splittings. However, the well-known degeneracy between $M_\tot$ and the current expansion rate, $H_0$ (e.g.~\cite{Planck:2018vyg}), leads to shifts in the posteriors of $H_0$ from $H_0 = 65.96_{-0.81}^{+1.3}\,\mathrm{km/s/Mpc}$ (IH) and $H_0 = 66.42_{-0.76}^{+1.3}\,\mathrm{km/s/Mpc}$ (NH) to $H_0 = 67.07_{-0.72}^{+1.2}\,\mathrm{km/s/Mpc}$ for the common DM approximation (all at 68\% C.L.). Therefore, imposing a physical hierarchy does have an impact on the inferred $H_0$, which -- taken at face value -- leads to a worsening of the Hubble tension \cite{Verde:2019ivm, DiValentino:2021izs, Kamionkowski:2022pkx, Poulin:2023lkg}.

\begin{figure}
    \centering
    \includegraphics[width=1\linewidth]{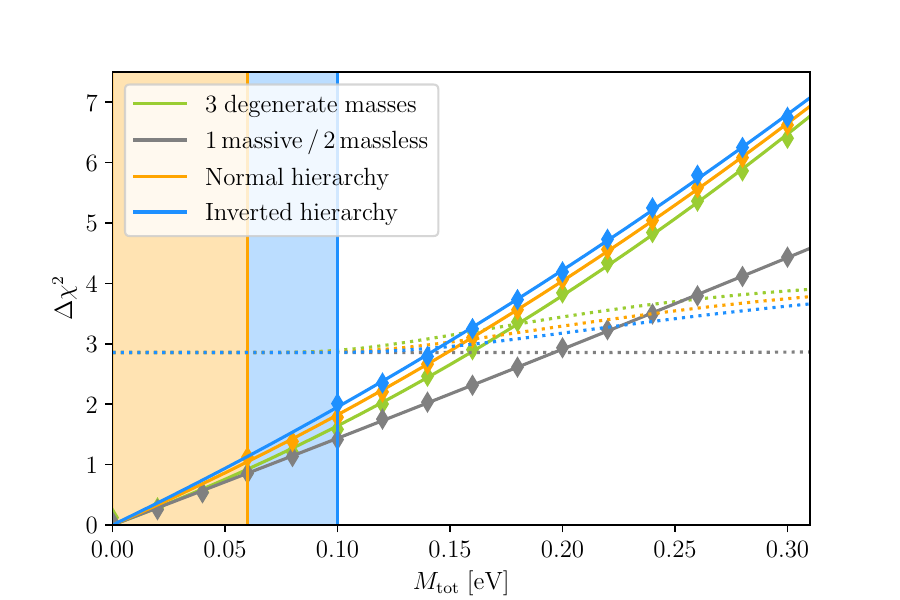}
    \includegraphics[width=1\linewidth]{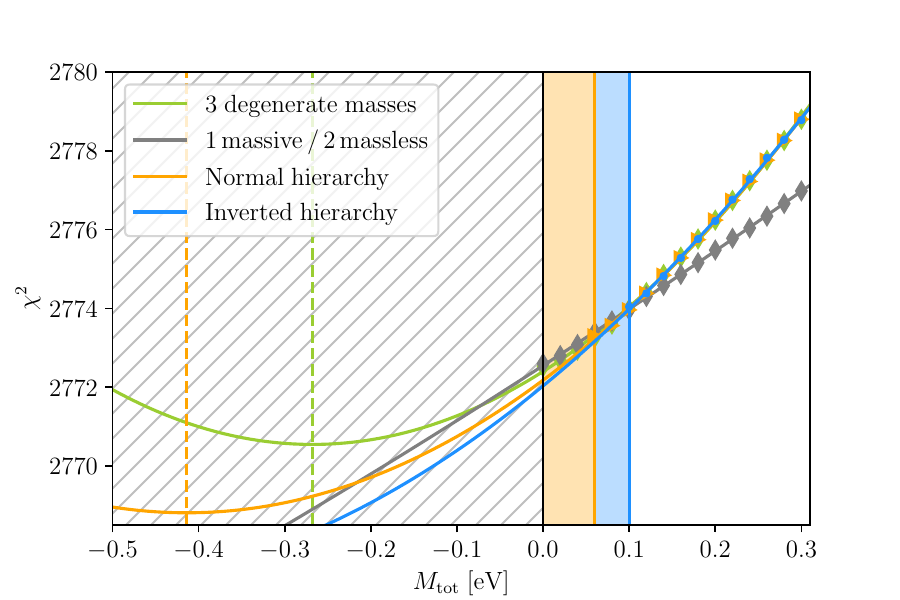}
    \caption{Profile likelihoods under \textbf{\Planck\ data} for different neutrino mass orderings. \textit{Top:} The intersection of the parabola fits (solid lines) with the respective dotted lines give the frequentist $95\%$ confidence intervals. The vertical orange and blue lines indicate the minimum allowed $M_\tot$ for NH and IH. \textit{Bottom:} Absolute $\chi^2$-values indicate that IH and NH fit the data marginally better than the 3 deg.-mass hierarchy and the 1 massive/2 massless scenario (for $M_\tot < 0.2\, \mathrm{eV}$). The minima of the extrapolated parabolas lie at $M_\tot < 0$ (vertical dashed lines).}
    \label{fig:PLs_Planck}
\end{figure}
The proximity of the inferred neutrino masses to the physical boundary $M_\tot > 0,\, 0.06,\, 0.1\, \mathrm{eV}$ for DM, NH and IH, respectively, can lead to subtleties in the analysis. Moreover, it is well known that neutrino mass constraints strongly depend on the assumed lower limit of the prior \cite{Simpson:2017qvj, Gariazzo:2018pei, Heavens:2018adv, Gariazzo:2023joe}. It is hence informative to compare the Bayesian credible intervals to frequentist confidence intervals from profile likelihoods. In Fig.~\ref{fig:PLs_Planck}, we compute profile likelihoods in $M_\tot$ under \Planck\ data assuming the four different neutrino mass orderings. The upper limits in $M_\tot$ are obtained at the intersection of the $\Delta\chi^2$ (solid lines) with the Neyman band (dotted lines in top panel of Fig.~\ref{fig:PLs_Planck}), where we assume a physical lower bound of $M_\tot > 0$.
The constraints are summarized in the right column of Tab.~\ref{tab:Planck_constraints}. As in the Bayesian case, the ``crude'' 1M approximation gives the loosest constraints ($M_\tot < 0.20\,\mathrm{eV}$, 1M), while the DM approximation gives slightly tighter constraints ($M_\tot < 0.18\,\mathrm{eV}$). In the frequentist setting, the two physically motivated hierarchies give the tightest constraints: $M_\tot < 0.16\,\mathrm{eV}$ (NH) and $M_\tot < 0.15\,\mathrm{eV}$ (IH). The reason for this becomes evident when looking at the absolute $\chi^2$ values obtained for the different neutrino mass orderings (bottom panel of Fig.~\ref{fig:PLs_Planck}). For a given fixed value of $M_\tot < 0.1\, \mathrm{eV}$,  DM fits \Planck\ data slightly better than the 1M approximation, while at $M_\tot \geq 0.1\, \mathrm{eV}$, the 1M approximation starts to fit the data better than the other mass hierarchies. This leads to a flatter profile likelihood for the 1M approximation and thus looser constraints. The profile likelihoods of DM, NH and IH, on the other hand, show excellent agreement, i.e.\ present a very similar fit to \Planck\ data.
The rather different behavior of 1M compared to the other neutrino mass orderings is not surprising given its stronger impact on the CMB power spectra and lensing as discussed in Sec.~\ref{sec:neutrino_physics} and \cite{Couchot:2017pvz, Archidiacono:2020dvx}.

With the profile likelihood, we can explore the impact of imposing a lower physical bound $M_\tot > 0.06\, \mathrm{eV}$ and $M_\tot > 0.1\, \mathrm{eV}$ motivated by the NH and IH, respectively. Imposing these limits as physical boundaries in the Feldman-Cousins construction loosens the bounds to $M_\tot < 0.23\,\mathrm{eV}$ and $M_\tot < 0.26\,\mathrm{eV}$ for the DM approximation. The constraints for the remaining neutrino hierarchies are similar and are quoted in parentheses in Tab.~\ref{tab:Planck_constraints}. 
This allows us to disentangle the effect of the lower limit from the effect of different predictions for the CMB and BAO observables in the different neutrino mass orderings. As can be seen from Tab.~\ref{tab:Planck_constraints}, the former has a stronger impact on the constraints than the latter. 

Extrapolating the profile likelihoods to negative $M_\tot$ illustrates the well-known preference of \Planck\ data for zero or even ``negative'' $M_\tot$ \cite{Planck:2013nga, Green:2024xbb, Naredo-Tuero:2024sgf, Herold:2024enb}. The minima of the extrapolated profile likelihoods lie deep in the un-physical negative regime for all hierarchies, between $-3.5\, \mathrm{eV}$ (1M) and $-0.27\, \mathrm{eV}$ (DM, dashed vertical lines in bottom panel of Fig.~\ref{fig:PLs_Planck}). 

Overall, the frequentist constraints are about $25$-$50\,\%$ tighter than the Bayesian constraints (when comparing NH and IH with matching lower bounds) confirming previous comparisons between Bayesian and frequentist neutrino mass constraint \cite{Naredo-Tuero:2024sgf, Herold:2024enb}. This in itself is not surprising due to the different interpretations and interval-construction methods in the Bayesian and frequentist frameworks, which can lead to discrepant constraints. These discrepancies can be expected to decrease with better data.


\subsection{\Planck\ and DESI}

\begin{figure*}
    \centering
    \includegraphics[width=1\linewidth]{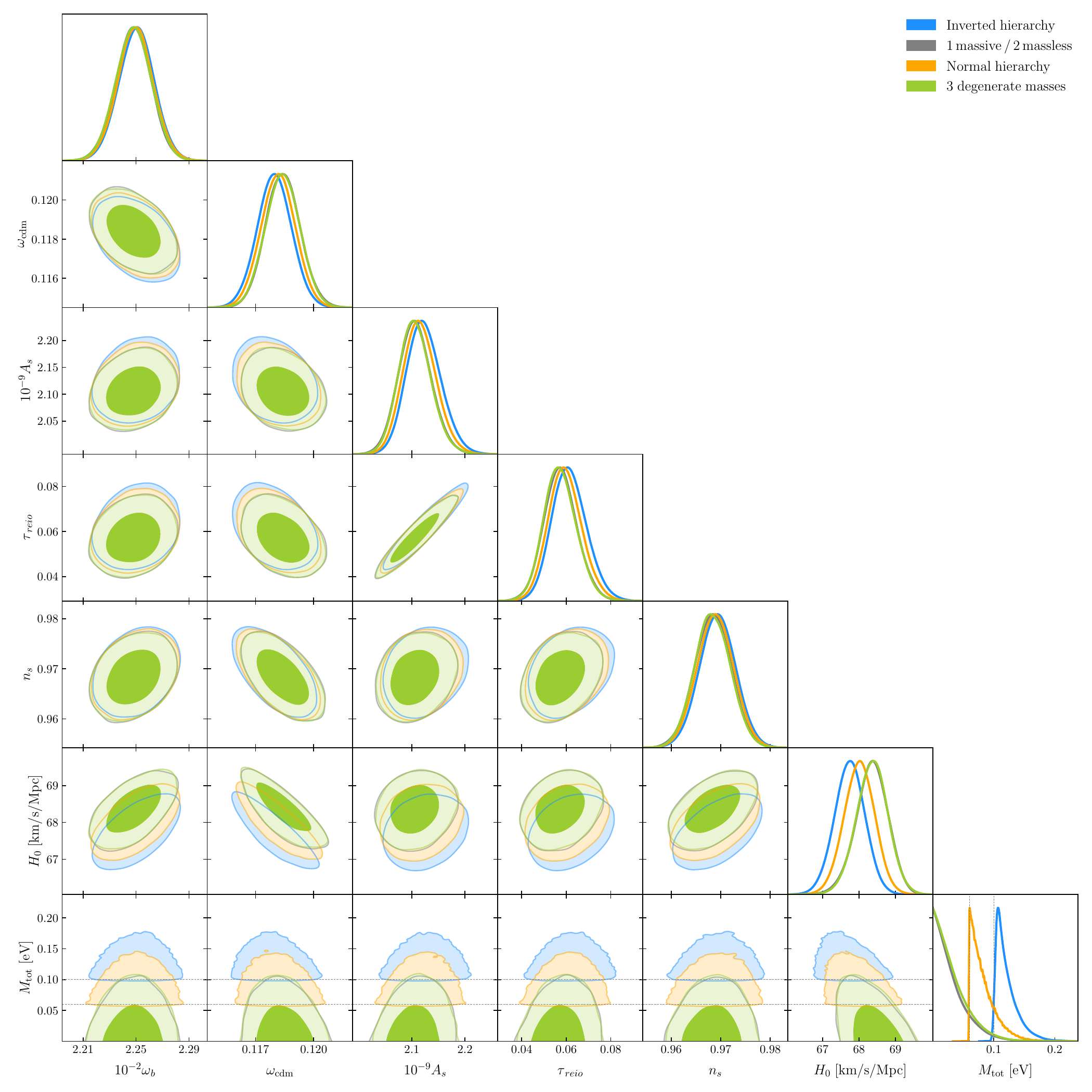}
    \caption{Posterior corner plot under \textbf{\Planck+DESI data} for different neutrino mass orderings. The 3-degenerate-masses and 1massive/2massless approximations give very similar constraints on $M_\tot$, which loosen when considering the physically motivated normal and inverted hierarchies due to the lower limit of the prior imposed by neutrino oscillation experiments (dotted lines in $M_\tot$ panels).}
    \label{fig:MCMC_PlanckDESI}
\end{figure*}
We now combine \Planck\ data with DESI BAO data. The constraints are summarized in Tab.~\ref{tab:PlanckDESI_constraints} and the posterior corner plot is shown in Fig.~\ref{fig:MCMC_PlanckDESI}. The inclusion of DESI BAO data tightens the Bayesian constraints for all hierarchies. The tightest constraints are obtained for the 1M approximation ($M_\tot < 0.083\,\mathrm{eV}$) and the DM approximation ($M_\tot < 0.086\,\mathrm{eV}$), which -- unlike for \Planck\ data -- give very similar results. Note that our upper limits from \Planck+DESI data are slightly higher than the ones in the DESI BAO baseline result, which assume the DM approximation ($M_\tot < 0.082\,\mathrm{eV}$, \cite{DESI:2024hhd}) since we do not include ACT lensing data.
Next we adopt the physically motivated NH and IH, which impose effective lower priors as described above (horizontal dotted lines in Fig.~\ref{fig:MCMC_PlanckDESI}). Similar to the case with \Planck\ data alone, the posteriors get cut off at higher values of $M_\tot$ leading to looser constraints: $M_\tot < 0.13\,\mathrm{eV}$ (NH) and  $M_\tot < 0.16\,\mathrm{eV}$ (IH). In App.~\ref{sec:prior}, we explicitly explore the impact of this lower limit by imposing the ``NH/IH prior'' in an analysis assuming the DM approximation. We find that the DM approximation with NH/IH priors agrees well with the full NH/IH analyses for \Planck+DESI data, reconfirming that the looser constraints from NH/IH are driven by the imposed lower limit. 

Similarly to the constraints under \Planck\ data alone, we find only negligible shifts of the posteriors of $\omega_b$, $\omega_\mathrm{cdm}$, $\tau$, $A_s$ and $n_s$. The inclusion of DESI BAO data partially breaks the $M_\tot$-$H_0$ degeneracy but a small trend of lower values of $H_0$ for higher values of $M_\tot$ is still present: from $H_0 = 67.75_{-0.44}^{+0.45}\,\mathrm{km/s/Mpc}$ (IH) and $H_0 = 68.01_{-0.44}^{+0.46}\,\mathrm{km/s/Mpc}$ (NH) to $H_0 = 68.36_{-0.43}^{+0.46}\,\mathrm{km/s/Mpc}$ (DM, all at 68\% C.L.). The inferred values of $H_0$ under \Planck+DESI data are slightly higher than under \Planck\ data alone due to the preference for slightly higher values of $H_0$ in DESI BAO data~\cite{DESI:2024mwx}.
\begin{table}[]
    \centering
    \begin{tabular}{c|c|c}
         Hierarchy & Bayesian & Frequentist (NH/IH lower limit)\\
         \hline
         DM        & $< 0.086\,\mathrm{eV}$ & $< 0.07\ (0.11/0.14)\,\mathrm{eV}$ \\
         1M        & $< 0.083\,\mathrm{eV}$ & $< 0.05\ (0.11/0.14)\,\mathrm{eV}$ \\
         NH        & $<  0.13\,\mathrm{eV}$ & $< 0.07\ (0.11/0.14)\,\mathrm{eV}$ \\
         IH        & $<  0.16\,\mathrm{eV}$ & $< 0.06\ (0.11/0.14)\,\mathrm{eV}$ \\
    \end{tabular}
    \caption{95\% upper limits on the sum of neutrino masses, $M_\tot$, under \textbf{\Planck+DESI data} for different neutrino mass hierarchies. For the frequentist interval construction, we assumed $M_\tot > 0$ ($M_\tot > 0.06$/$0.1\,\mathrm{eV}$) as a physical lower boundary (see text).}
    \label{tab:PlanckDESI_constraints}
\end{table}

\begin{figure}
    \centering
    \includegraphics[width=1\linewidth]{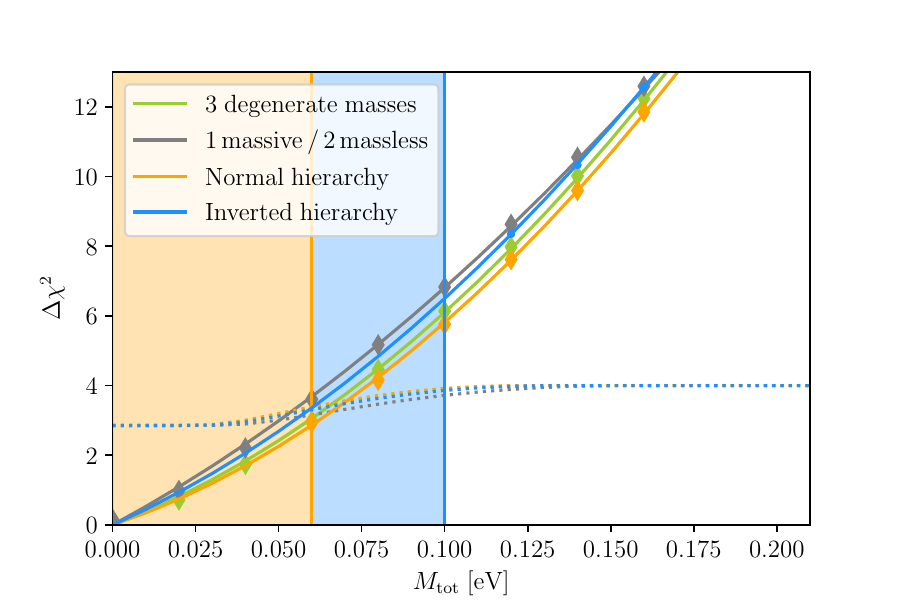}
    \includegraphics[width=1\linewidth]{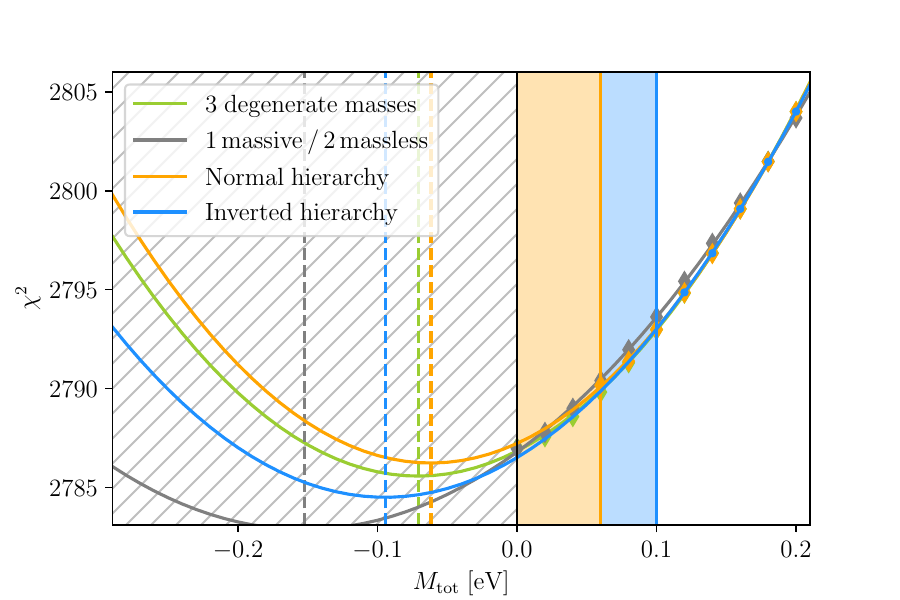}
    \caption{Profile likelihoods under \textbf{\Planck+DESI data} for different neutrino mass orderings. \textit{Top:} The intersection of the parabola fits (solid lines) with the respective dotted lines give the frequentist $95\%$ confidence intervals. The vertical orange and blue lines indicate the minimum allowed $M_\tot$ for NH and IH. \textit{Bottom:} Absolute $\chi^2$-values indicate that all mass orderings present a very similar fit to the data. The minima of the extrapolated parabolas lie at $M_\tot < 0$ (vertical dashed lines).}
    \label{fig:PLs_PlanckDESI}
\end{figure}
To gain a deeper understanding of the tight upper limits from \Planck+DESI data, we compare the Bayesian constraints to frequentist constraints obtained with profile likelihoods (right column of Tab.~\ref{tab:PlanckDESI_constraints} and Fig.~\ref{fig:PLs_PlanckDESI}). All assumed hierarchies give  tight constraints, where the tightest is obtained with the 1M approximation ($M_\tot < 0.05\,\mathrm{eV}$). Slightly looser constraints are obtained for IH ($M_\tot < 0.06\,\mathrm{eV}$) and for DM and NH (both $M_\tot < 0.07\,\mathrm{eV}$). Note that the upper limits for the IH are \textit{below} the minimum allowed mass of $0.1\,\mathrm{eV}$ in the IH; thus, we will consider physically motivated lower bounds below. The different neutrino mass orderings give rise to  similar profile likelihoods, where the difference in the profile likelihoods and in the upper limits is similar to the level of the numerical noise in the minimizations.\footnote{At this precision, the inferred upper limits depend sensitively on the $\Delta\chi^2$-value around $0.05-0.07\,\mathrm{eV}$ and the width of the extrapolated parabola, which introduces some uncertainty due to the proximity to the physical boundary in $M_\tot = 0$.} 

As discussed in Sec.~\ref{sec:neutrino_physics}, the inclusion of BAO data partially breaks the geometrical degeneracy in the CMB between $M_\nu$, $\Omega_m$ and $H_0$. This leaves less wiggle room in the cosmological parameters to adapt the fit to CMB data and can explain the similar quality of fit to \Planck+DESI data for all neutrino mass orderings. The impact of the different orderings on the geometry and background can explain the small differences in the four scenarios.

To disentangle the effect of imposing a lower physical bound $M_\tot > 0.06\, \mathrm{eV}$ (NH) and $M_\tot > 0.1\, \mathrm{eV}$ (IH) from the difference in the physical modeling, we conduct the Feldman-Cousins construction with these boundaries.  For all neutrino models, this yields identical constraints quoted in parentheses in Tab.~\ref{tab:PlanckDESI_constraints}: $M_\tot < 0.11\,\mathrm{eV}$ (``NH lower bound'') and $M_\tot < 0.14\,\mathrm{eV}$ (``IH lower bound''). Hence, imposing the lower limit dominates over the physical modeling differences in the different neutrino mass orderings. 

The goodness of fit quantified by the absolute $\chi^2$-values under \Planck+DESI data (bottom panel of Fig.~\ref{fig:PLs_PlanckDESI}) makes it evident that all neutrino mass orderings fit the data similarly well, as the profiles agree for $0<M_\tot<0.2\,\mathrm{eV}$, with the 1M approximation presenting a slightly worse fit. This reaffirms that the Bayesian constraints for NH and IH give looser constraints than the DM approximation due to the effective prior imposed by neutrino oscillations experiments. 

Extrapolating the profiles into the un-physical negative regime (hatched area) reaffirms the preference for zero or ``negative'' $M_\tot$ for all neutrino mass orderings. The minima of the extrapolated parabolas lie between $-0.15\, \mathrm{eV}$ (1M) and $-0.06\, \mathrm{eV}$ (NH), closer to zero than for \Planck\ data alone but still more than $2\sigma$ away from $0.06\, \mathrm{eV}$, the minimum allowed $M_\tot$ in the NH. However, previous studies have found that this preference for ``negative'' masses diminishes for different data sets, for example for alternative \Planck\ pipelines. Refs.~\cite{Allali:2024aiv, Naredo-Tuero:2024sgf, RoyChoudhury:2024wri} find that the neutrino mass constraints from \Planck\ PR3 data (used in this work) are affected by the ``lensing anomaly'' (e.g.\ \cite{Calabrese:2008rt, Addison:2023fqc}) and replacing it with \Planck\ PR4 and the \texttt{HiLLiPoP}/\texttt{LoLLiPoP} \cite{Tristram:2023haj} or \texttt{CamSpec} \cite{Rosenberg:2022sdy} likelihoods leads to relaxed constraints, while \cite{DESI:2024hhd} find comparable results with all \Planck\ pipelines when combining with DESI full-shape clustering data. Moreover, allowing for more complex cosmological models than $\Lambda$CDM (assumed here), e.g.\ dynamical dark energy models, loosens the constraints \cite{Hannestad:2005gj, Vagnozzi:2018jhn, RoyChoudhury:2018vnm, Elbers:2024sha, Naredo-Tuero:2024sgf, RoyChoudhury:2024wri, DESI:2024hhd}.

Similar to the \Planck-only analysis, the frequentist constraints from \Planck+DESI data are about 10-40\% tighter (when comparing NH and IH with matching lower bounds). Note that these constraints on $M_\tot$ are considerably tighter than frequentist constraints \cite{Naredo-Tuero:2024sgf, Herold:2024enb} from \Planck\ \cite{Planck:2018vyg} and previous BAO data from 6dF \cite{Beutler:2011hx}, SDSS \cite{Ross:2014qpa} and BOSS \cite{BOSS:2016wmc}, for which frequentist and Bayesian constraints give similar results. As discussed above, discrepancies between frequentist and Bayesian approaches are not unexpected due to the different interpretations. It can be expected that these discrepancies decrease with the advancement of cosmological data sets. Regardless, the tight upper limits from \Planck+DESI data in the frequentist framework for all assumed neutrino mass orderings reaffirms the Bayesian picture of constraints that come very close to the minimum allowed masses from neutrino oscillation experiments.


\section{Conclusions}
\label{sec:conclusions}

In this work, we explored the impact of different neutrino mass orderings on the inferred mass constraints from \Planck\ and \Planck+DESI data.

For \Planck\ data, we find that the DM approximation gives tighter constraints than the physically motivated NH and IH (Fig~\ref{fig:MCMC_Planck}). This loosening of the constraints can be attributed to the lower bound imposed by neutrino oscillation experiments for NH and IH (Eq.~\ref{eq:min_mass}), confirming previous results. This is corroborated by the frequentist profile likelihood analysis, which shows that the DM approximation fits the data similarly well as NH and IH (Fig.~\ref{fig:PLs_Planck}). The approximation with one massive and two massless neutrinos (1M), however, presents a worse fit than the other neutrino mass orderings and gives looser Bayesian and frequentist constraints. Despite \cite{Hannestad:2003ye, Lesgourgues:2004ps, Crotty:2004gm, Lesgourgues:2012uu, Lesgourgues:2013} pointing out that the 1M approximation leads to significant changes in the background and perturbations compared to NH/IH/DM, the 1M approximation is often set as a default configuration in samplers like \texttt{MontePython} and \texttt{Cobaya} \cite{Torrado:2020dgo}. Our exploration confirms that the 1M approximation should be avoided and replaced by the DM approximation in parameter inference.

For \Planck+DESI data, the DM and 1M approximations give tighter Bayesian constraints than the physically motivated NH and IH (Fig.~\ref{fig:MCMC_PlanckDESI}). The loosening of the constraints for NH and IH can be attributed to the imposed lower limits in NH and IH, respectively, which we explore explicitly by imposing the same lower limits for the DM approximation (Fig.~\ref{fig:MCMC_PlanckDESI_prior}). The profile likelihood analysis shows that all neutrino models fit the data similarly well (Fig.~\ref{fig:PLs_PlanckDESI}), thus indicating that \Planck+DESI data is not sensitive to the details of the neutrino mass ordering.

For both data set combinations, we extrapolate the profile likelihoods into the un-physical negative regime, finding that the extrapolated minima of the parabolas lie at $M_\tot < 0$, re-affirming the preference for zero or ``negative'' $M_\tot$ \cite{Green:2024xbb, Noriega:2024lzo, Naredo-Tuero:2024sgf, Elbers:2024sha}. 
We recover the well-know degeneracy between $M_\tot$ and $H_0$, which leads to lower values of $H_0$ for higher values of $M_\tot$ as obtained with NH and IH, which in turn leads to a worsening of the Hubble tension. 

The frequentist upper limits are consistently tighter than the Bayesian constraints. While this is not unexpected due to the different interpretations in the two statistical approaches, it will be interesting to see if this discrepancy disappears for future data. Regardless, the frequentist constraints confirm the persistently tighter neutrino mass constraints from \Planck+DESI data. While this may eventually lead to a tension between cosmology and terrestrial neutrino experiments, previous studies have shown that other cosmological data sets, e.g.\ alternative \Planck\ likelihoods, give rise to considerably loosened constraints \cite{Allali:2024aiv, Naredo-Tuero:2024sgf, RoyChoudhury:2024wri}. 
Moreover, we want to emphasize that neutrino mass constraints from cosmology are model dependent. While in this work we assumed the $\Lambda$CDM model throughout, more complex models can lead to relaxed constraints \cite{Hannestad:2005gj, Vagnozzi:2018jhn, RoyChoudhury:2018vnm, Elbers:2024sha, Naredo-Tuero:2024sgf, RoyChoudhury:2024wri, DESI:2024hhd}.

Overall, our work shows that the DM hierarchy presents a good approximation for the physically motivated NH and IH even when including DESI BAO data \textit{if} the same lower limits are imposed in the analysis.


\vspace{4mm}
\textit{Data availability:} The data and code that support the findings of this article are openly available \cite{Planck:link, DESI:link, Herold:link}.

\vspace{4mm}
\textit{Acknowledgements: } We are grateful for stimulating discussions with Graeme Addison, Andrea Cozzumbo, Elisa Ferreira, Wayne Hu, Tanvi Karwal, Julien Lesgourgues and Kushal Lodha. LH was supported by a \textit{William H. Miller} fellowship. MK was supported by NSF Grant No.\ 2412361, NASA Grant No.\ 80NSSC24K1226, and the John Templeton Foundation. This work was carried out at the Advanced Research Computing at Hopkins (ARCH) core facility ({\tt arch.jhu.edu}), which is supported by the National Science Foundation (NSF) grant number OAC1920103.


\appendix

\section{Impact of the lower limit of $M_\tot$ imposed by neutrino oscillation experiments}
\label{sec:prior}

\begin{figure*}
    \centering
    \includegraphics[width=1\linewidth]{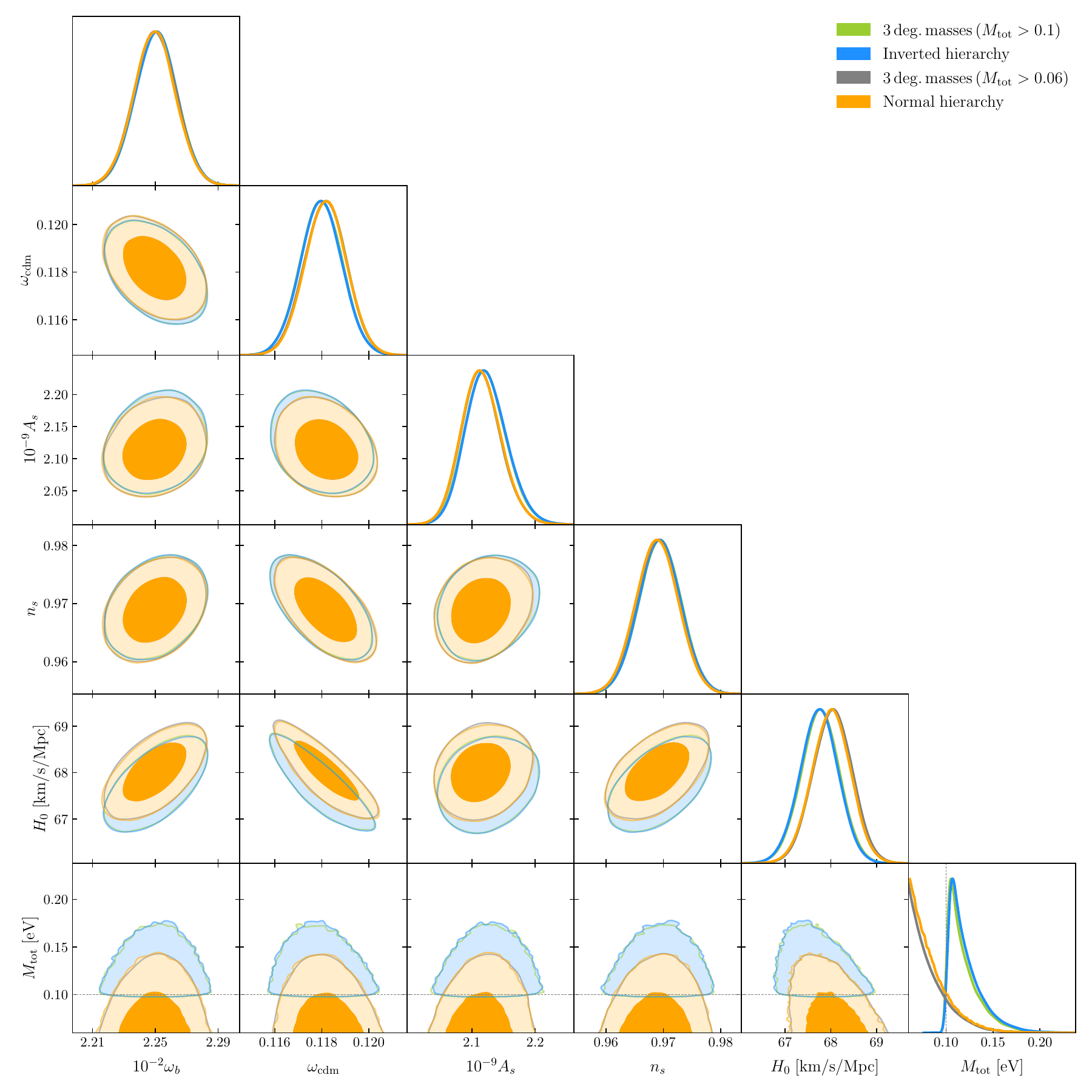}
    \caption{Posterior corner plot under \textbf{\Planck+DESI data} for NH and IH as well as the DM approximation with lower bounds $M_\tot >0.06\,\mathrm{eV}$ (NH) and $M_\tot >0.1\,\mathrm{eV}$ (IH). This illustrates that the lower limit imposed is the dominant difference between the DM approximation and NH/IH in the Bayesian analysis.}
    \label{fig:MCMC_PlanckDESI_prior}
\end{figure*}

The lower limit inferred from neutrino oscillation data, which imposes a flat prior with lower limit $M_\tot > 0.06\,\mathrm{eV}$ (``NH prior'') and $M_\tot > 0.1\,\mathrm{eV}$ (``IH prior''), makes it difficult to directly compare Bayesian constraints from NH/IH to the DM approximation, where for the latter we adopted the common lower limit of the prior $M_\tot > 0$. A straightforward comparison can be achieved by imposing the same lower limits as in NH/IH to the common DM approximation (as was also explored in \cite{DESI:2024mwx}). We show the resulting posteriors under \Planck+DESI data in Fig.~\ref{fig:MCMC_PlanckDESI_prior}. As expected, the constraints of the DM approximation with NH prior ($M_\tot < 0.12\,\mathrm{eV}$) agree with the analysis adopting the full NH mass ordering and the DM approximation with IH prior ($M_\tot < 0.16\,\mathrm{eV}$) agrees well with the full IH mass ordering. 

\bibliography{main}

\end{document}